\documentclass[12pt,a4paper]{article}
\usepackage{amssymb}
\usepackage{graphicx}
\begin{document}

\title{Can tetraneutron exist from theoretical point of view?}
\author{I.V. Simenog$^{a,b}$ , B.E. Grinyuk$^{a}$ , Yu.M. Bidasyuk$^{b}$ \\
\\
$^{a}$\textit{Bogolyubov Institute for Theoretical Physics,} \\
\textit{Nat. Acad. of Sci. of Ukraine,} \textit{Kyiv 03143, Ukraine} \\
(e-mail: isimenog@bitp.kiev.ua; bgrinyuk@bitp.kiev.ua)\\
$^{b}$ \textit{Taras Shevchenko Kyiv National University,}\\
\textit{\ Kyiv 03127, Ukraine}}
\maketitle

\begin{abstract}
{\small A theoretical possibility is shown for the bound state of a
tetraneutron to exist in the case of the proposed neutron-neutron potentials
in the singlet state with two attractive wells separated by a repulsive
barrier. The anomalous behaviours are revealed for the calculated size,
density distribution, and pair correlation functions of a hypothetical
tetraneutron.}
\end{abstract}


\section{Introduction}

The mysterious fact of the experimental registration of
tetraneutrons \cite {1,2} in a reaction with loosely bound
radioactive $^{14}$Be renewed the attention to theoretical
attempts of understanding the problem of hypothetical bound
neutron systems. The experiment contradicts the rather old
estimates showing the impossibility to form a bound state of a few
neutrons interacting by the standard nuclear forces (see review
\cite{3}). Recent attempts to study this problem using more modern
methods of calculation of four-particle systems \cite{4,5,6} also
indicate the impossibility of binding the four-neutron system
without adding some exotic many-particle interaction potentials.
Moreover, the absence of resonances in a four-neutron system with
standard potentials was shown in \cite{7}. The experimental search
for resonances in $^{4}n$ and $^{3}n$ systems \cite{8} using other
nuclear reactions had also no success.

In the present paper, we study the theoretical problem of the possible
existence of a tetraneutron by developing the precise methods of calculation
of loosely bound states of four Fermi particles. We propose a special idea
of constructing the neutron-neutron potentials allowing the bound
tetraneutron to exist and simultaneously describing the standard low-energy
neutron-neutron data.

\section{Basic equations}

To study the properties of the four-neutron system in the state with zero
spin ($S=0$) and orbital moment ($L=0$) under assumption of the central
pairwise neutron-neutron interaction potentials, we solve the following
Schr\"{o}dinger equation for one spatial component of the wave function:

\begin{eqnarray}
&&\{\sum_{i=1}^{4}\frac{\mathbf{p}_{i}^{2}}{2m}+\frac{1}{2}%
\sum_{i>j=1}^{4}\left( V_{s}^{+}(r_{ij})+V_{t}^{-}(r_{ij})\right) +\frac{1}{2%
}\sum_{(ij)\neq (14),(23)}(-1)^{i+j}\left(
V_{s}^{+}(r_{ij})-V_{t}^{-}(r_{ij})\right)  \label{E1} \\
&&-\frac{1}{2}\;\sum_{(ij)\neq (12),(34)}(-1)^{i+j}\left(
V_{s}^{+}(r_{ij})-V_{t}^{-}(r_{ij})\right) \hat{P}_{23}\}\Phi
=E\Phi .  \nonumber
\end{eqnarray}
The total antisymmetric wave function of the four-neutron system is
expressed in terms of the corresponding spin and spatial components as

\begin{equation}
\Psi ^{a}(1,2,3,4)=\frac{1}{\sqrt{2}}\left( \Phi ^{\prime }(\mathbf{r}_{1},%
\mathbf{r}_{2},\mathbf{r}_{3},\mathbf{r}_{4})\xi ^{\prime \prime }-\Phi
^{\prime \prime }(\mathbf{r}_{1},\mathbf{r}_{2},\mathbf{r}_{3},\mathbf{r}%
_{4})\xi ^{\prime }\right) ,\;  \label{E2}
\end{equation}
where the antisymmetric $\Phi ^{\prime }$ and symmetric $\Phi ^{\prime
\prime }$ (with respect to permutations ($1\rightleftarrows 2$) and ($%
3\rightleftarrows 4$)) spatial components are

\bigskip
\begin{eqnarray}
&&\Phi ^{\prime }(\mathbf{r}_{1},\mathbf{r}_{2},\mathbf{r}_{3},\mathbf{r}%
_{4})\equiv \Phi (\mathbf{r}_{1},\mathbf{r}_{2},\mathbf{r}_{3},\mathbf{r}%
_{4}),  \label{E3} \\
&&\Phi ^{\prime \prime }(\mathbf{r}_{1},\mathbf{r}_{2},\mathbf{r}_{3},%
\mathbf{r}_{4})\equiv \frac{1}{\sqrt{3}}\left( 2\Phi (\mathbf{r}_{1},\mathbf{%
r}_{2},\mathbf{r}_{3},\mathbf{r}_{4})-\Phi (\mathbf{r}_{1},\mathbf{r}_{2},%
\mathbf{r}_{3},\mathbf{r}_{4})\right) ,  \nonumber
\end{eqnarray}
which corresponds to the Young scheme [2,2]. In Eq. (\ref{E1}),
$\hat{P}_{23}$
is the permutation operator of spatial coordinates, $V_{s}^{+}(r_{ij})$ and $%
V_{t}^{-}(r_{ij})$ are, respectively, the singlet interaction
potential in even states and the triplet one in odd states. The
bound states of four neutrons are studied by solving the
Schr\"{o}dinger equation (\ref{E1}) for various nuclear potentials
taken in the form of a superposition of Gaussian functions, by
using the well-known variational method with the translationally
invariant Gaussian basis antisymmetrized with respect to the
permutations of particles ($1\rightleftarrows 2$) and
($3\rightleftarrows 4$),

\begin{equation}
\Phi (\mathbf{r}_{1},\mathbf{r}_{2},\mathbf{r}_{3},\mathbf{r}_{4})=\hat{A}%
\sum_{k=1}^{N}C_{k}\exp \left( -\sum_{i>j=1}^{4}u_{ij}^{k}r_{ij}^{2}\right)
,\;  \label{E4}
\end{equation}
where $N$ is the basis dimension, and $r_{ij}\equiv \left| \mathbf{r}_{i}-%
\mathbf{r}_{j}\right| $. This basis and the special schemes necessary to
optimize the nonlinear variational parameters $u_{ij}^{k}$ enable us to
carry on the calculations of loosely bound states with desired high accuracy.

\section{Spinless interaction model}

First, consider the simplest case of potentials independent of spin ( i.e. $%
V_{t}^{-}(r_{ij})=V_{s}^{+}(r_{ij})$) where the overestimated
attraction in the triplet state may only promote the binding of
the four-neutron system. We study the $^{4}n$ bound state
appearance conditions varying the coupling constant of potentials
of different forms (further, we use the
dimensionless units: $V(r)=\frac{\hslash ^{2}}{mr_{0}^{2}}U(\frac{r}{r_{0}}%
)\equiv \frac{\hslash ^{2}}{mr_{0}^{2}}gu(r),$ where $r_{0}$ is the radius
of interaction, and $g$ is the coupling constant; $\hbar ^{2}/m=41.4425$ MeV$%
\cdot $fm$^{2}$ for neutrons). For the potential with one Gaussian
function, $U(r)=-g\exp (-r^{2})$, a bound tetraneutron $^{4}n$
exists below the decay threshold ($4\rightarrow 2+2$) only for
$g\geq g_{cr}(4)=3.911$, which is $1.46$ times greater than the
critical two-particle coupling constant $g_{cr}(2)=2.684$.
We notice that the reliable calculations need basis (\ref{E4}) to be about $%
150$ functions with the optimization of nonlinear parameters. Note that a
trineutron, for the same potential, can be bound below the decay threshold ($%
3\rightarrow 2+1$) only for $g\geq 3.3g_{cr}(2)$. From the
qualitative point of view, the similar conditions of the
tetraneutron bound state appearance take place for other
traditional purely attractive potentials. Thus, under the
condition that a dineutron is unbound, there is no possibility,
because of the Pauli principle, for a tetraneutron (nothing to say
of a trineutron) to form a bound state with traditional attractive
potentials. An analogous conclusion can be drawn in the case of
common potentials with repulsion at short distances between
neutrons. For example, for the widely used Volkov potential, one
has $k\equiv g_{cr}(4)/g_{cr}(2)=1.44$. Moreover, the attempt to
find a better ratio $k$ by varying the parameters of the
two-component potentials with attraction and short-range repulsion
led us only to a potential $U(r)=g\left( 1.5\exp
(-(r/0.9)^{2})-\exp (-r^{2})\right) $ giving rise to $k$ about
1.27. We can assume that it is impossible to form a bound system
$^{4}n$ also for other standard interaction potentials with
attraction and short-range repulsion if $^{2}n$ is unbound, which
is in agreement with the recent calculations \cite{4,5,6}.

In principle, a possibility for a bound tetraneutron to exist,
under the condition of an unbound $^{2}n$, can be realized with
some exotic pairwise interaction potentials having two regions of
attraction separated by a repulsive barrier. An external
attractive potential well of greater radius is necessary, first of
all, to fit the experimental low-energy two-neutron scattering
parameters in the singlet state, and it has to be in the typical
range of nuclear forces of about or greater $1.5$ fm. Fitting the
singlet interaction potential, we use the following low-energy
neutron-neutron scattering parameters: the scattering length
$a_{s(nn)}=-18.9$ fm and effective radius $r_{0s(nn)}=2.75$ fm.
The internal attractive potential well of smaller radius is
important for binding the $^{4}n$ system, while the repulsive
barrier between the attractive wells makes the two regimes of
attraction somewhat independent. In a tetraneutron, the number of
pairs of particles in the singlet state, as well as that in the
triplet one, equals three, the Pauli principle reveals itself only
in the triplet state, and the internal potential well acting in
the singlet state plays the main role in binding the $^{4}n$
system. A class of potentials with two
attractive wells of different radii, which give rise to the bound state of $%
^{4}n$ under the assumption $V_{t}^{-}(r_{ij})=V_{s}^{+}(r_{ij})$, is rather
wide. We have a number of potentials in the form of a superposition of three
or four Gaussian functions. One of the optimal variant is the four-component
neutron-neutron singlet potential

\begin{equation}
U_{s}^{+}(r)=g\left\{ 0.43\exp \left( -\left( r/0.6\right) ^{2}\right) -\exp
\left( -r^{2}\right) +1.085\exp \left( -\left( r/1.3\right) ^{2}\right)
-0.42\exp \left( -\left( r/1.5\right) ^{2}\right) \right\} ,\;  \label{E5}
\end{equation}
where the distance is measured in units of $r_{0}=0.488519$ fm. At $g=$ $%
g_{exper}=322.40$, potential (\ref{E5}) reproduces the
experimental values of $a_{s(nn)}$, $r_{0s(nn)}$, and the commonly
used recommended singlet neutron-neutron phase shift up to the
energies $E_{\rm lab}\thickapprox 80 $ MeV. Note that there are no
direct measurements of the neutron-neutron phase shifts.

\begin{figure}[tbp]
\centering
\includegraphics[width=6in]{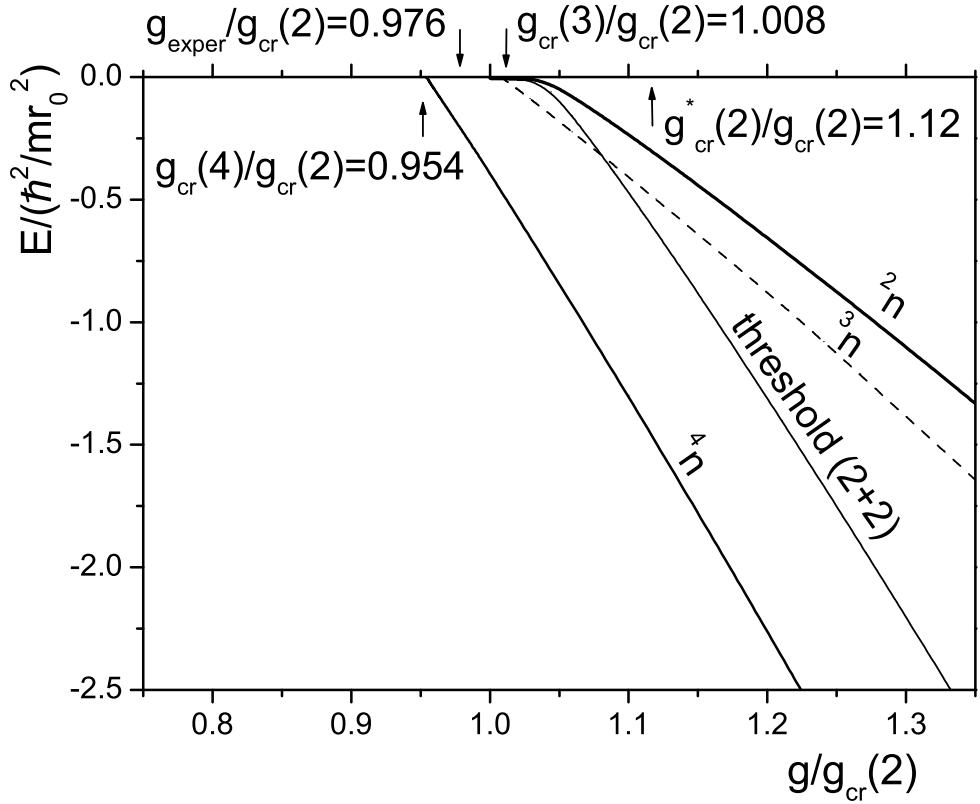}
\caption{Tetraneutron energy dependence on the coupling constant
of potential (\ref {E5}) acting both in the singlet and triplet
states ($r_{0}=0.488519$ fm is the radius of the interaction
potential (\ref {E5})).} \label{fig1}
\end{figure}

Fig.~\ref{fig1} shows the dependence of the tetraneutron energy on
the coupling
constant $g$ of potential (\ref{E5}) in the ''spinless'' case $%
U_{t}^{-}(r_{ij})=U_{s}^{+}(r_{ij})$. Note that the decay threshold of $^{4}n
$ into $2+2$ (as well as the decay threshold of $^{3}n$ into $2+1$) as a
function of the coupling constant has two regimes of behaviour. The first
regime of a rather weak binding of $^{2}n$ at $g\longrightarrow g_{cr}(2)$
takes place due to the presence of the attraction of greater radius in
potential (\ref{E5}), and the second one with the almost linear dependence
of the threshold in a wide range of coupling constants is present due to the
attraction of smaller radius. A repulsive barrier between the attractive
wells contributes to the sharpness of changing the two regimes of the
threshold behaviour. Note also that the excited two-particle $S$-state lies
anomalously close to the ground state at $g_{cr}^{\ast }(2)/g_{cr}(2)=1.12$,
which is caused to a great extent by the presence of two almost independent
attractive wells in potential (\ref{E5}). It is essentially important that,
in variational calculations, the $^{4}n$ system is bound already at $g\geq
g_{cr}(4)=315.2=0.954g_{cr}(2)$, where $^{2}n$ is still unbound ($g\leq
g_{cr}(2)=330.42$). Moreover, at the coupling constant $%
g=g_{exper}=322.40=0.976g_{cr}(2)$, where potential (\ref{E5})
reproduces the experimental low-energy neutron-neutron parameters,
the $^{4}n$ system is already bound. At the same time, a
trineutron with the considered potential is not allowed to be
bound since the ratio $g_{cr}(3)/g_{cr}(2)=1.008$ is greater than
$1$, although being close to it. Notice the fact that the $^{4}n $
energy dependence on $g$ looks like almost a straight line
parallel to the energy threshold ($2+2$) dependence in a wide
interval of coupling constants ($g\gtrsim 1.1$), and this line is
rather close to the ($2+2$) threshold. This fact indicates that,
in this region of $g$, a tetraneutron exists due
to the presence of the internal potential well of smaller radius, and the $%
^{4}n$ state is of the two-dineutron cluster nature. A similar consideration
concerns $^{3}n$ as well: in a wide interval of coupling constants, it is
the cluster state ($2+1$) with the essential role of the attractive well of
smaller radius. This is confirmed also by the approximate relation $%
E_{^{4}n}-2E_{^{2}n}\thickapprox 2\left( E_{^{3}n}-E_{^{2}n}\right) $. By
the way, we constructed some other variants of potentials $U_{s}^{+}(r)$,
for example,

\[
U_{s}^{+}(r)=g\left\{ 0.315\exp \left( -\left( r/0.5\right) ^{2}\right)
-\exp \left( -r^{2}\right) +1.278\exp \left( -\left( r/1.31\right)
^{2}\right) -0.54\exp \left( -\left( r/1.5\right) ^{2}\right) \right\} ,\;
\]
which yields $g_{cr}(4)/g_{cr}(2)=0.9525$ for $^{4}n$ to exist and
reproduces the low-energy parameters of $n-n$ scattering at $%
g_{exper}/g_{cr}(2)=0.9935$ ($g_{cr}(2)=187.5$, and interaction radius $%
r_{0}=1.2608$ fm). Even a trineutron could exist for this potential upon the
unbound $^{2}n$ due to the ratio $g_{cr}(3)/g_{cr}(2)=0.9564$ being less
than unity, but the more strict condition $g_{cr}(3)<g_{exper}$ is not
valid. In addition, the $n-n$ singlet phase shift for this potential becomes
too large already at rather low energies. We have found no variant of the
potential obeying the condition $g_{cr}(3)<g_{exper}$ simultaneously with
giving a reasonable phase shift at low energies for $^{3}n$ to exist in a
bound state.

It should be noted that it is necessary to carry on variational
calculations with special schemes of optimization of basis
(\ref{E4}) in order to obtain the above results for $^{4}n$ with
reliable accuracy. In particular, we used about 220 functions with
the optimization of the basis for this purpose. This is caused by
both the complicated antisymmetrized four-neutron wave function of
the near-threshold state and the potential containing essentially
different components.

\section{Realistic case}

Now consider more realistic models of the triplet interaction potential $%
U_{t}^{-}(r)$, when the conditions for the existence of the bound state of a
tetraneutron are somewhat less appropriate. If we put $U_{t}^{-}(r)$ to be
zero in Eq. (\ref{E1}), we get an unbound tetraneutron within the proposed
class of singlet potentials, because only a half of 6 pairs interacts in
this case. Thus, it is necessary to have some additional attraction in odd
orbital states to bind $^{4}n$. On the other hand, the commonly recommended
phase shifts of the scattering in odd orbital states are rather negative
corresponding to the effective repulsion. It appears that there exists a
class of triplet potentials which together with the singlet potential (5)
can bind the $^{4}n$ system and are repulsive with the exception of the
typical nuclear distances of about $1.5-2$ fm, where they reveal some
attraction correlated with the external attractive potential well of the
singlet potential. Such a potential (in the same dimensionless units, as
potential (\ref{E5})) may have the form

\begin{equation}
U_{t}^{-}(r)=g_{t}\left\{ 2.212\exp \left( -\left( r/2\right) ^{2}\right)
-2.334\exp \left( -\left( r/3\right) ^{2}\right) +\exp \left( -\left(
r/4\right) ^{2}\right) \right\} \;  \label{E6}
\end{equation}
with $g_{t}=14$. This potential has negative phase shift in the
$P$-state, although with non-monotone dependence. Potential
(\ref{E6}) together with the singlet one (\ref{E5}) result in the
bound state of a tetraneutron with the binding energy
$B(^{4}n)\gtrsim 0.5$ MeV (this value is the variational
estimation with the use of about 400 Gaussian functions). A more
accurate calculation needs much greater efforts mainly because of
the complicated structure of potentials and the many-component
antisymmetrized wave function of four particles. In addition, the
ultimate result for the binding energy is a few orders of
magnitude lesser than the contributions of the kinetic or
potential energies calculated separately with rather good
accuracy, and these contributions almost cancel each other having
opposite signs. The proposed phenomenological potentials are
constructed only to demonstrate the possibility for a tetraneutron
to exist in a bound state. Moreover, one can easily change the
binding energy of $^{4}n$ in a wide range (from zero to dozens of
MeV) by changing slightly potential (\ref{E6}) or potential
(\ref{E5}). There are some reasons to assume that if these
potentials with repulsive barriers are changed in such a way that
they should not allow a tetraneutron to be bound, they may result
in resonances in the system of four neutrons.

\begin{figure}[tbp]
\centering
\includegraphics[width=6in]{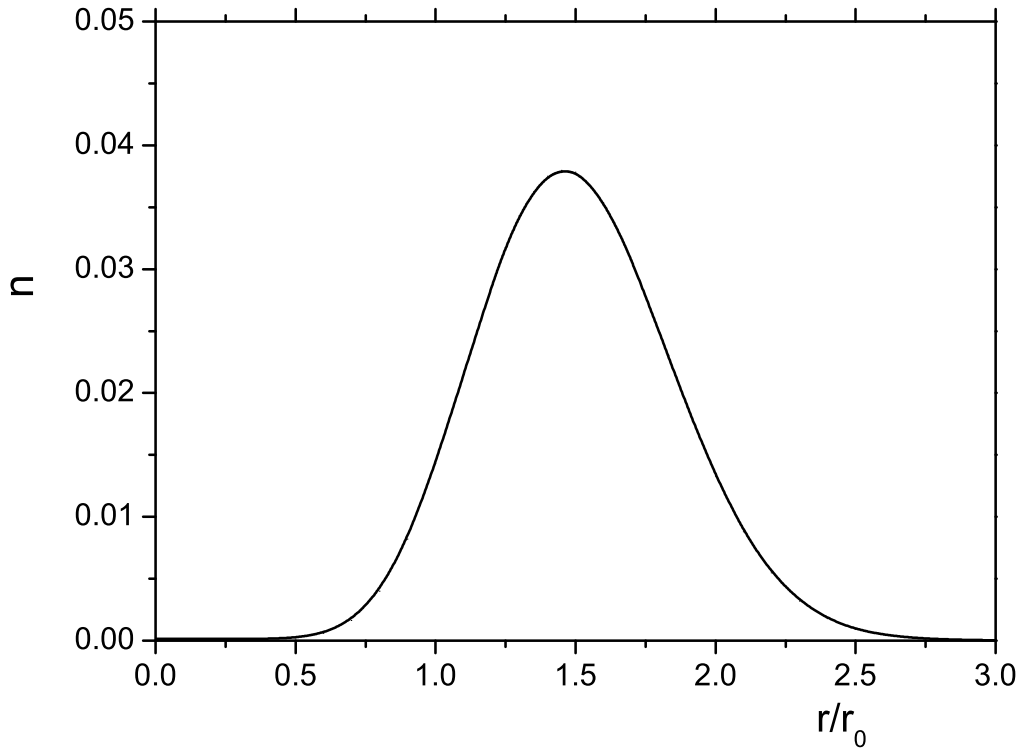}
\caption{Profile of the density distribution $n(r)$ of neutrons of
the hypothetical tetraneutron ($r_{0}$ is the same as in
Fig.~\ref{fig1}). } \label{fig2}
\end{figure}

Consider the main structure functions of the hypothetical
tetraneutron. Note that the structure functions can be calculated
much more accurately using basis (4) of a lesser dimension than
that used in the calculation of the energy. Fig.~\ref{fig2}
presents the one-particle density distribution of $^{4}n$
(normalized as $\int n(r)d\mathbf{r}=1$ ) versus the dimensionless
distance, for potentials (\ref{E5}), (\ref{E6}). Due to the Pauli
principle, the density distribution has essential minimum at short
distances, i.e. the tetraneutron is a ''bubble'' system with the
almost Gaussian near-surface distribution of neutrons. The
tetraneutron has anomalously small (in nuclear scale) r.m.s.,
$\left\langle r^{2}\right\rangle ^{1/2}=1.704r_{0}=0.83$ fm, which
is caused by the attraction well of smaller radius in potential
(\ref{E5}). Changing the singlet potential (\ref{E5}), one can
increase the above value mainly due to an increase of the radius
$r_{0}$. But, in any case, the size of a tetraneutron, in spite of
its extremely small binding energy, will be less or about the size
of an $\alpha $-particle.

\begin{figure}[tbp]
\centering
\includegraphics[width=6in]{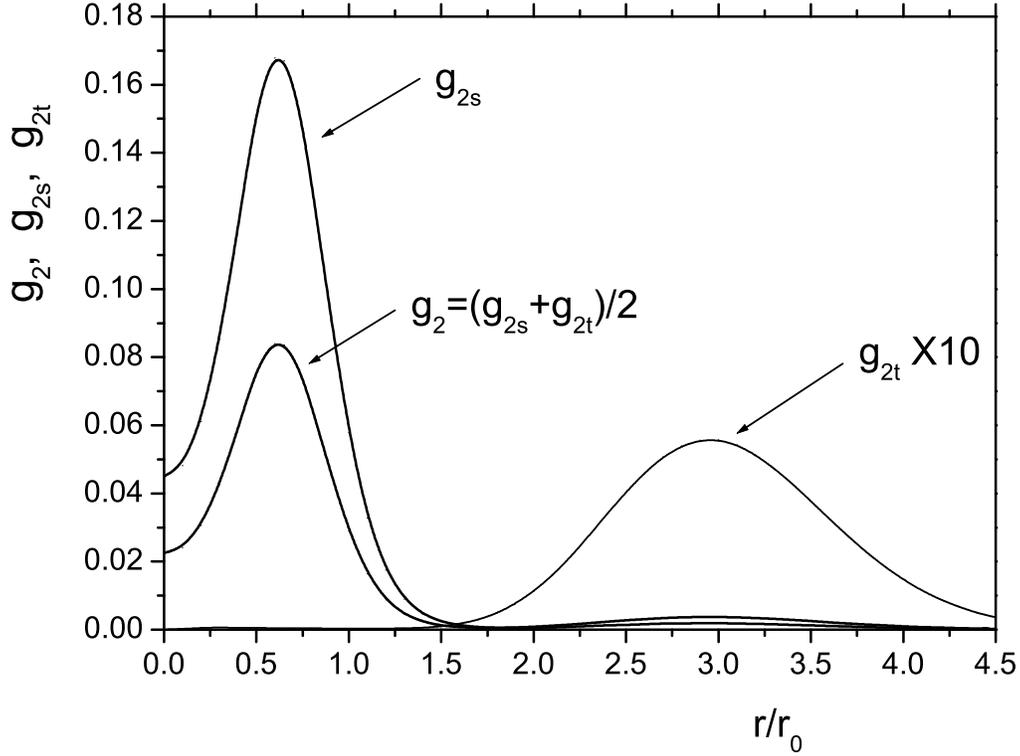}
\caption{Profiles of the singlet $g_{2s}(r)$, triplet $g_{2t}(r)$, and total $%
g_{2}(r)=\frac{1}{2}\left( g_{2s}(r)+g_{2t}(r)\right) $ pair
correlation functions of the hypothetical bound $^{4}n$ system
($r_{0}$ is the same as in Fig.~\ref{fig1}).} \label{fig3}
\end{figure}

Fig.~\ref{fig3} depicts the singlet $g_{2s}(r)$ and triplet
$g_{2t}(r)$ pair correlation functions, which reflect, to a great
extent, the behaviour of the corresponding potentials. The singlet
correlation function $g_{2s}(r)$ has significant maximum in the
region of internal short-range attraction of the singlet
potential, since the Pauli principle does not reveal itself in the
singlet state. Some decrease of $g_{2s}(r)$ at very short
distances is caused by the presence of short-range repulsion in
potential (\ref{E5}), and it should not be present if the
repulsion were absent. The secondary maximum is present in
$g_{2s}(r)$ due to the existence of the external attractive
potential well of $V_{s}^{+}(r)$. In the triplet state, the
repulsion at short distances makes a small contribution into the
energy because of the Pauli principle, and the maximum of
$g_{2t}(r)$ is located in the attractive area of the triplet
potential. That is why, the contribution of the triplet potential
to the energy of $^{4}n$ is negative, and a tetraneutron could not
be bound without the contribution of this comparatively small
effective attraction. The short-range attraction in the singlet
state plays the main role in binding the $^{4}n$ system. This is
confirmed by calculations of the average singlet and triplet
potential energy contributions,

\[
\left\langle V\right\rangle =3\left\{ \int
V_{s}^{+}(r)g_{2s}(r)d\mathbf{r}+\int
V_{t}^{-}(r)g_{2t}(r)d\mathbf{r}\right\} \equiv \left\langle
V_{s}^{+}\right\rangle +\left\langle V_{t}^{-}\right\rangle ,\;
\]
where $\left\langle V_{s}^{+}\right\rangle =-1296.7$ MeV and
$\left\langle
V_{t}^{-}\right\rangle =-158.7$ MeV, which together with the kinetic energy $%
\left\langle \hat{K}\right\rangle =1454.9$ MeV result in the
negative energy of the system $E_{^{4}n}\lesssim -0.5$ MeV
indicated above.

The total correlation function $g_{2}(r)=\frac{1}{2}\left(
g_{2s}(r)+g_{2t}(r)\right) $ reflects the average neutron pair correlations
and has the main maximum at short distances and the secondary one in the
region of attraction in the triplet state (where the singlet potential also
has an external attractive well).

\section{Conclusions}

To summarize, we note the following. 1) A tetraneutron can exist
in the bound state if one assumes that the interaction potential
in the $n-n$ singlet state has two attractive wells separated by a
repulsive barrier. Unfortunately, as a result, we get an
anomalously high maximum in the singlet scattering phase shift
$\delta _{s}\thickapprox 160^{0}$ at the energies of neutrons of
the order of \ $100-150$ MeV. The problem of constructing the
potential, which binds $^{4}n$ and gives no anomalous maximum
mentioned above, is to be further studied. Maybe, a combination of
pairwise potentials with some small intercluster ones can improve
the situation. The nature of the assumed exotic short-range
attraction is to be discussed as well. 2) Strange though it may
seem, few-nucleon systems are essentially underbound with the
$n-n$ potentials (\ref{E5}), (\ref{E6}) used together with the
standard $n-p$ ones. In particular, with the Minnesota potential
used as the $n-p$ interaction, one has the binding energies of
about $6.2$ MeV for $^{3}$H, and about $23.4$ MeV for $^{4}$He.
The calculated few-nucleon values may be in agreement with the
experimental data under the condition that the fitting of the
potentials is carried out taking into account the concordance of
the regimes of attraction of all the potentials. 3) The
hypothetical tetraneutron has abnormal small size and, at the same
time, small binding energy, and this may serve as an additional
criterion for the identification of such systems. It is
interesting to study the probability of the tetraneutron
''presence'' in $^{14}$Be nuclei. 4) Potentials (\ref{E5}) and
(\ref{E6}) satisfy the saturation conditions necessary for the
stability of the neutron matter. Interesting and nontrivial
questions arise concerning heavier multineutron systems with such
potentials. In particular, a system of 8 neutrons can be a more
stable system than $^{4}n$ both because of the magic number of
particles, and because the former may be like two tetraneutron
clusters.

\section{Acknowledgements}

The authors thank to Prof. A.G. Zagorodny for the discussions and support of
this work.

\end{document}